\documentclass[twoside]{article}
\input{stefmac.sty}
\usepackage{graphics,color,latexsym,amssymb,amsmath,amsfonts,verbatim,alltt,url}
\usepackage[dvips]{graphicx}
by 2 \advance \topmargin -2in \leftmargin 210mm \advance \leftmargin
-\textwidth \divide \leftmargin by 2 \advance \leftmargin -.9in
\oddsidemargin \leftmargin \evensidemargin \leftmargin
\begin{document}

\thispagestyle{plain} \setcounter{section} {0} \setcounter{footnote} {0} %
\setcounter{equation} {0} \thispagestyle{empty}
\setcounter{page}{1}

\noindent\textbf{ROMANIAN JOURNAL OF INFORMATION }\newline
\textbf{SCIENCE AND TECHNOLOGY}\newline \noindent Volume
\textbf{12}, Number 2, 2009, 265–279

\markboth{Sofronia, Popa, Stefanescu}{Undecidability Results for Finite Interactive Systems}\
\vspace{1.5cm}

\begin{center}
{\LARGE \bf Undecidability Results for Finite Interactive Systems}\bigskip

{\large Alexandru Sofronia$^a$ and Alexandru Popa$^b$ and Gheorghe Stefanescu$^{c,}$\foo{On leave from the
    Faculty of Mathematics and Computer Science, University of Bucharest.}}\bigskip

$^a$ Faculty of Mathematics and Computer Science, University of Bucharest\\
\vspace{2pt} E-mail: {\tt alexandrusofronia@yahoo.com} \\ [9pt]

$^b$ Department of Computer Science, University of Bristol\\
\vspace{2pt} E-mail: {\tt popa@cs.bris.ac.uk} \\ [9pt]

$^c$ Department of Computer Science, University of Illinois at Urbana-Champaign\\
\vspace{2pt} E-mail: {\tt ghstef@yahoo.com} \\ [9pt]

\end{center}
\bigskip

\begin{quote}{\small \qquad 

{\bf Abstract.}~ A new approach to the design of massively parallel and interactive programming languages has
been recently proposed using rv-systems (interactive systems with registers and voices) and Agapia
programming. In this paper we present a few theoretical results on FISs (finite interactive systems), the
underlying mechanism used for specifying control and interaction in these systems. First, we give a proof for
the undecidability of the emptiness problem for FISs, by reduction to the Post Correspondence Problem. Next,
we use the construction in this proof to get other undecidability results, e.g., for the accessibility of a
transition in a FIS, or for the finiteness of the language recognized by a FIS. Finally, we present a simple
proof of the equivalence between FISs and tile systems, making explicit that they precisely capture
recognizable two-dimensional languages.

}
\end{quote}

\vspace{18pt}{\large \bf 1.~Introduction
}\vspace{12pt}

A new approach to the design of massively parallel and interactive programming languages has been recently
proposed. The approach focuses on Agapia \cite{dr-st08b,pss07}, a programming language paradigm based on
classical register machines and space-time duality \cite{stefanescu00,ste06}. Agapia extends usual programming
languages with coordination features, being a step forward to the integration of coordination features (as in
Klaim \cite{nic+98}, Reo \cite{arb04}, Orc \cite{mi-co07}, etc.) into practical programming languages.  A few
distinctive features of Agapia are: high-level modularity, including a structured approach to interaction
based on name-free processes; simple operational and relational semantics based on grids and scenarios
(enriched two-dimensional words); invariance with respect to space-time duality.

Agapia language uses (i) complex spatial and temporal data for inerfaces, (ii) modules over usual programming
languages, and (iii) temporal, spatial, or spatio-temporal while-programs for coordination. In Agapia v0.1 one
can write programs for open processes located at various sites and having their temporal windows of adequate
reaction to the environment. It naturally supports process migration, structured interaction, and deployment
of modules on heterogeneous machines. Agapia can be seen as an extension of usual procedural or functional
programming languages (used in the basic modules), for instance may be developed on top of languages as C,
Java, Scheme, etc.

The theoretical foundation of Agapia is strongly related to the theory of two-dimensional languages
\cite{gi-re97,grst96,lmn98,la-si97b}. It is based on FISs (finite interactive systems) \cite{ste02,ste06},
abstract mechanisms for specifying control and interaction in interactive systems. FISs can be used to
recognize two-dimensional languages, and, in this respect, FISs are equivalent to tile systems \cite{gi-re97},
existential monadic second order logic \cite{grst96}, or other equivalent presentations of regular (or
recognizable) two-dimensional languages \cite{gi-re97,lmn98}. However, they come equipped with abstract
``states'' and ``interaction classes,'' which, by instantiation, may be used to design interactive programs.

The present paper contains a few results on FISs. First, it presents a proof for the undecidability of the
emptiness problem for FISs by reduction to the Post Correspondence Problem. Next, the construction in this
proof is used to show the accessibility problem for FISs (i.e., whether, for a specific transition, there is
an accepting scenario for a two-dimensional word using that transition) is undecidable. Finally, the paper
includes a simple and direct proof of the equivalence between FISs and tile systems, emphasizing the
conceptual differences between these two equivalent presentations of recognizable two-dimensional language; as
a byproduct, this gives another (this time, indirect) proof of the undecidability of the emptiness problem for
FISs via the undecidability of a similar problem for tile systems. A conference version of the paper has
appeared in \cite{sps08}.

\vspace{18pt}{\large \bf 2.~Preliminaries
}\vspace{12pt}

\paragraph{Grids and scenarios}

A {\em grid} (also called a {\em two-dimensional word}) is a {\em rectangular} two-dimensional area filled in
with letters of a given alphabet.  The columns in a grid correspond to processes, the top-to-bottom order
describing their progress in time. The left-to-right order corresponds to process interaction in a {\em
  nonblocking message passing discipline}: a process sends a message to the right, then it resumes its
execution.

\begin{figure}[bth]\begin{center}\hspace*{-.5cm}\begin{tabular}{c@{\hsp\hsp}cc}
\raisebox{.7cm}{\tdwbis{small}{aabbabb\\abbcdbb\\bbabbca\\ccccaaa}}
& \raisebox{.9cm}{$\tdwbis{small}{\ 1\ 1\ 1\ \tdret AaBbBbB\tdret \ 2\ 1\
1\ \tdret AcAaBbB\tdret \ 2\ 2\ 1\ \tdret AcAcAaB\tdret 2\ 2\ 2\ }$}
& $F_1={}$ \raisebox{-.8cm}{\includegraphics[scale=.10]{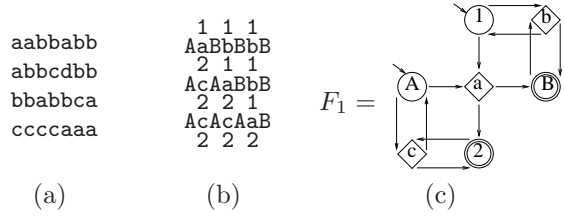}}\svsp\\
(a)&(b)&(c)
\end{tabular}\nvsp\end{center}
\caption{A grid (a), an abstract scenario (b), and a FIS (c).}\label{f-gscen}
\end{figure}

A {\em scenario} is a grid enriched with data around each letter. The data may have various interpretations:
they either represent control/interaction information, or current values of the variables, or both. In this
paper, we only use abstract scenarios of the first type resulting from accepting runs in finite interactive
systems. A grid is presented in Fig.~\ref{f-gscen}(a) and an abstract scenario in Fig.~\ref{f-gscen}(b).

We use the following notation for operations on grids: $\vcomp$ and $\vstar$ denote vertical composition and
iteration, while $\hcomp$ and $\hstar$ denote the horizontal composition and iteration.

\paragraph{Finite interactive systems} 
 
A {\em finite interactive system} (shortly FIS) is a finite hyper-graph with two types of vertices and one
type of (hyper) edges: the first type of vertices is for {\em states} (labeled by numbers), the second is for
{\em classes} (labeled by capital letters) and the edges/transitions are labeled by letters denoting the atoms
of the grids; each transition has two incoming arrows (one from a class and the other from a state), and two
outgoing arrows (one to a class and the other to a state). Some classes/states may be {\em initial} (indicated
by small incoming arrows) or {\em final} (indicated by double circles); see, e.g., \cite{ste02,ste06}.

For the {\em parsing procedure}, given a FIS $F$ and a grid $w$, insert initial states/classes at the
north/west border of $w$ and parse the grid completing the scenario according to the FIS transitions; if the
grid is fully parsed and the south/east border contains final states/classes only, then the grid $w$ is
recognized by $F$.  The {\em language} of $F$ is the set of its recognized grids. 

Let $F_1$ be the FIS graphically represented as in Fig.~\ref{f-gscen}(c). It is equivalently represented
specifying its transitions $\scruce{1}{A}{a}{B}{2},\hsp \scruce{1}{B}{b}{B}{1}, \hsp \scruce{2}{A}{c}{A}{2}$
and pointing out that $A,1$ are initial and $B,2$ final. A parsing for $\tdwbis{footnotesize}{abb\tdret
  cab\tdret cca}$ is
$$\begin{array}{cccccc} 
\tdwbis{footnotesize}{\ 1\ 1\ 1\ \tdret Aa\ b\ b\ \tdret \ \ \ \ \ \ \ \tdret Ac\ a\ b\ \tdret \ \ \ \ \ \ \
\tdret Ac\ c\ a\ \tdret \ \ \ \ \ \ \ }
& \tdwbis{footnotesize}{\ 1\ 1\ 1\ \tdret AaBb\ b\ \tdret \ 2\ \ \ \ \ \tdret Ac\ a\ b\ \tdret \ \ \ \ \ \ \
\tdret Ac\ c\ a\ \tdret \ \ \ \ \ \ \ }
& \dots
& \tdwbis{footnotesize}{\ 1\ 1\ 1\ \tdret AaBbBbB\tdret \ 2\ 1\ 1\ \tdret AcAaBbB\tdret \ 2\ 2\ 1\ \tdret
AcAcAaB\tdret 2\ 2\ 2\ }\end{array}$$ 

\paragraph{Post Correspondence Problem}

Let $x = (x_1,\ldots,x_n)$ and $y = (y_1,\ldots,y_n)$ be two lists of nonempty words from an alphabet
$\Sigma$, with at least two letters.  The {\it Post Correspondence Problem (PCP)} is to decide whether or not
there exist $i_1,\ldots,i_k$ where $k\geq 1$ and\foo{$\overline{p,q}$ denotes the set $\{p,p+1,\dots,q\}$}
$1\leq i_j\leq n, \forall j\in \overline{1,k}$ such that $x_{i_1} \ldots x_{i_k} = y_{i_1} \ldots y_{i_k}.$ It
is known that PCP is undecidable \cite{Post47} (if $|\Sigma|\geq2$).

We use this result to prove the emptiness problem for FISs is undecidabile.

\vspace{18pt}{\large \bf 3.~The emptiness problem and the finiteness problem for FISs
}\vspace{12pt}

Let $x = (x_1,\ldots,x_n)$ and $y = (y_1,\ldots,y_n)$ be an instance of the PCP, labeled $PCP(x,y)$. We
construct a finite interactive system $S$ which accepts a language $L=L(S)$ such that: $L$ is finite iff it is
empty iff $PCP(x,y)$ has no solution. Let $|w|$ denote the length of the string $w\in \Sigma$.\vsp

The idea of the construction is as follows. The FIS associated to the PCP instance allows to parse grids where
the first two rows contain a proposed solution for the PCP. More precisely, it contains sequences of $x_i$'s
and $y_j$'s (with a single letter in a cell) such that their product is equal. The next rows check if the
chosen indices are equal, hence whether or not one gets a solution for the PCP.

The states and the classes of S are:

{\bf States}
\begin{itemize}
\item $s$
\item $a(i,j)$ $\forall i\in \overline{1,n}\;\;\forall j\in \overline{1,|x_i|}$ 
\item $c(i,j)$ $\forall i\in \overline{0,n}\;\;\forall j\in \overline{0,n}$
\end{itemize}

{\bf Classes}
\begin{itemize}
\item $A$
\item $B(i,j)$ $\forall i\in \overline{1,n}\;\;\forall j\in \overline{1,|x_i|-1}$ 
\item $C(i,k)$ $\forall i\in \overline{1,n}\;\;\forall k\in \overline{1,|y_i|-1}$
\item $M(i,j,k)$ $\forall i\in \overline{1,n}\;\;\forall j\in \overline{0,|x_i|}\;\;\forall k\in \overline{0,|y_i|}$
\end{itemize}

$S$ has a unique initial state - $s$, a unique initial class - $A$, a unique final state - $c(0,0)$ and $n+1$
final classes - $A$ and $M(i,0,0)$ for each $i\in \overline{1,n}$. To simplify the definition of the
transitions we use an extended notation $A=B(i,0)=B(i,|x_i|)=C(i,0)=C(i,|y_i|)$ $\forall i\in \overline{1,n}$.

Let $x_i = x_i^1\cdot x_i^2\cdots x_i^{|x_i|}$, where $x^k_i$'s are the letters of the word $x_i$, $1\leq
i\leq n$ and use a similar notation for $y_i$'s. The alphabet of the FIS $S$ also contains a special symbol
$\$$ such that $\$\not\in \Sigma$.

We define the {\bf transitions} of $S$ as follows:
\renewcommand{\labelenumi}{(\Roman{enumi})}
\begin{enumerate}
\item \cross{s}{B(i,j-1)}{x_i^{j}}{\addspace{.35cm}{B(i,j)}}{\addspace{1.25cm}{a(i,j)}} $\forall i\in
\overline{1,n}\;\;\forall j\in \overline{1,|x_i|}$
\item \cross{a(j,g)}{C(i,k-1)}{y_i^{k}}{\addspace{.35cm}{C(i,k)}}{\addspace{1.25cm}{c(j,i)}} $\forall i\in
\overline{1,n}\;\;\forall k\in \overline{1,|y_i|}$ iff $x_j^{g}=y_i^{k}$, where $j\in \overline{1,n}$ and
$g\in \overline{1,|x_j|}$
\item \cross{c(0,0)}{\addspace{.75cm}{A}}{\$}{\addspace{1.5cm}{A}}{\addspace{.35cm}{c(0,0)}}
\item \cross{c(i,i)}{\addspace{.75cm}{A}}{\$}{M(i,|x_i|-1,|y_i|-1)}{\addspace{.35cm}{c(0,0)}} $\forall i\in
\overline{1,n}$
\item \cross{c(i,0)}{\addspace{.75cm}{A}}{\$}{\addspace{.35cm}{M(i,|x_i|-1,|y_i|)}}{\addspace{.35cm}{c(0,0)}}
$\forall i\in \overline{1,n}$
\item \cross{c(0,i)}{\addspace{.75cm}{A}}{\$}{\addspace{.3cm}{M(i,|x_i|,|y_i|-1)}}{\addspace{.35cm}{c(0,0)}}
$\forall i\in \overline{1,n}$
\item \cross{c(j_1,j_2)}{M(i,k_1,k_2)}{\$}{\addspace{.8cm}{M(i,r_1,r_2)}}{\addspace{.05cm}{c(m_1,m_2)}} 

where $i\in \overline{1,n}$, $j_1,j_2,m_1,m_2\in \overline{0,n}$, $k_1,r_1\in \overline{0,|x_i|}$, $k_2,r_2\in
\overline{0,|y_i|}$ and
$k_1$ satisfies exactly one of the following conditions
\begin{itemize}
\item $k_1=0,\;m_1=j_1,\;r_1=k_1.$
\item $k_1>0,\;j_1=i,\;m_1=0,\;r_1=k_1-1.$
\item $k_1>0,\;j_1=0,\;m_1=0,\;r_1=k_1.$
\end{itemize}
and $k_2$ satisfies exactly one of the following conditions
\begin{itemize}
\item $k_2=0,\;m_2=j_2,\;r_2=k_2.$
\item $k_2>0,\;j_2=i,\;m_2=0,\;r_2=k_2-1.$
\item $k_2>0,\;j_2=0,\;m_2=0,\;r_2=k_2.$
\end{itemize}
\end{enumerate}

The following lemmas reveal the behavior of this FIS and the role of the transitions. Recall that the grids
below are parsed from left to right and from top to bottom.

Note that equalities of indexes in the following lemmas must be interpreted as equalities of lists over
$\{1,\ldots,n\}$ rather than weaker string equalities. We use a simplified notation $i^n$ instead of
$(i,\ldots,i)$, where $i$ appears $n$ times in a row.
\vsp

\renewcommand{\labelenumi}{(\roman{enumi})}
\noi{\bf Lemma 1}
Let $w$ be a grid with $m$ lines and $q$ columns. If there exists a successful running for $w$ with respect to
the border conditions\foo{I.e., using on the borders the specified sequences $b_n,b_w,b_s,
  b_e$ of initial/final states/classes} $b_n,b_w,b_s, b_e$, then:
\begin {enumerate}
\item $m\geq 3$, $q\geq 1$, $b_n \in \{s\}^{\star}$, $b_w \in \{A\}^{\star}$, $b_s\in \{c(0,0)\}^{\star} $.
\item There exist $k\geq 1$, $r\geq 1$, $i_1,\ldots,i_k$ and $j_1,\ldots,j_r$ where $1\leq i_l\leq n \;
  \forall l\in \overline{1,k}$ and $1\leq j_l\leq n \; \forall l\in \overline{1,r}$ such that $x_{i_1} \cdots
  x_{i_k}$ and $y_{j_1} \cdots y_{j_r}$ are the first two lines of $w$ (as strings) and $x_{i_1} \cdots
  x_{i_k} = y_{j_1} \cdots y_{j_r}$.
\item The southern border of the second line of $w$ is $c(a_1,b_1), \ldots , c(a_q,b_q)$ where\\ $(a_1,\ldots,
  a_q)= (i_1^{|x_{i_1}|}, \ldots,i_k^{|x_{i_k}|})$ and $(b_1,\ldots, b_q) = (j_1^{|y_{j_1}|}, \ldots
  j_r^{|y_{j_r}|})$ (as lists over $\{1,\ldots,n\}$)
\item The first two lines of $w$ are parsed using type $(I)$ and type $(II)$ transitions, and no other lines
  can be parsed using type $(I)$ or type $(II)$ transitions. In particular, all the lines of $w$, except the
  first two, are composed of $\$$.
\vsp
\end{enumerate}

\bproof\vsp

{\it (i)} Since $s$ is not a final state, it follows that $q\geq 1$ and $m\geq 1$. The conditions for the
border are trivial from uniqueness of initial and final states and initial classes.

Recall that the only final state of $S$ is $c(0,0)$. A successful run of $w$ has at least three transitions
from a northern border $s$ to a southern border $c(0,0)$: a type $(I)$ transition, followed by a type $(II)$
transition and, as $i,j>0$ in type $(II)$ transitions, a type $(IV)$ or $(VII)$ transition. Therefore, in
order to be accepted, $w$ has to have at least three lines.\vsp

{\it (ii)} First, since $A$ is the only initial class and $s$ the only initial state, only type $(I)$
transitions can be used to parse the first letter of $w$. Therefore this letter must be the first letter of
some word $x_i$. Note that this argument can be applied whenever we have $A$ on the western border and $s$ on
the northern border.

All the transitions on the first line of $w$ have $s$ as the northern border and therefore are type $(I)$
transitions. Such transitions contain only letters from the words in the list $x$. Note that these transitions
can be connected horizontally only if the corresponding letters are adjacent in some $x_i$ (this is the role
played by the $B(i,j)$ classes) or are the final letter of some $x_i$ and the first letter of some $x_k$ (in
which case, the class in the middle is $A = B(i,|x_i|)=B(k,0)$).

It follows that the words from the list $x$ that appear in the first line of $w$ cannot be truncated (except
for possibly the rightmost one). Since type $(I)$ transitions cannot have any $M(i,0,0)$ as a eastern border,
the eastern border of the first line of $w$ must be $A$, the only final class remaining. Therefore, the first
line of $w$ is some $x_{i_1} \cdots x_{i_k}$ where $k\geq 1$ (as $q\geq 1$).

The northern border of the second line of $w$ is made of $a(i,j)$ states, therefore only type $(II)$
transitions can be used for parsing. Since $C(i,k)$ classes prevent truncation, similar to the $B(i,j)$
classes, a similar argument shows that the second line of $w$ is $y_{j_1} \cdots y_{j_r}$ for some $r\geq 1$,
and it is parsed using only type $(II)$ transitions.

A type $(I)$ transition accepting a letter $x_i^{k}$ has on the southern border the state $a(i,k)$. The first
index encodes the position of the word $x_i$ in the list $x$ and the second index encodes the position $k$ in
which the letter $x_i^{k}$ appears in $x_i$. The condition $x_j^{g}=y_i^{k}$ in the definition of type $(II)$
transitions forces the letters in the first and second line to be identical. Therefore, $x_{i_1} \cdots
x_{i_k} = y_{j_1} \cdots y_{j_r}$.\vsp

{\it (iii)} From {\it (ii)} it follows that the southern border of the first line has the following format
$a(i_1,1) a(i_1,2) \dots a(i_1,|x_{i_1}|) \dots a(i_k,1) \dots a(i_k,|x_{i_k}|)$. As type $(II)$ transitions
copy the first index of the northern border $a(j,g)$ to the first index of the southern border $c(j,i)$, the
southern border of line 2 (or northern border of line 3) can be written $c(a_1,b_1), \ldots , c(a_q,b_q)$,
and, by extracting the first indexes, $(a_1,\ldots , a_q) = (i_1^{|x_{i_1}|}, \ldots ,i_k^{|x_{i_k}|})$. The
second index of the state $c(j,i)$ in a type $(II)$ transition is equal to the position of the word $y_i$ in
the list $y$, so from {\it (ii)} we get $(b_1,\ldots ,b_q) = (j_1^{|y_{j_1}|}, \ldots ,j_r^{|y_{j_r}|})$.\vsp

{\it (iv)} Recall that the first line of $w$ is parsed using only type $(I)$ transitions and the second line
using only type $(II)$ transitions. To conclude the proof, note that the northern border of line 3 is made of
$c(*,*)$ states, type $(III)-(VII)$ transitions also produce $c(*,*)$ states on the southern border and type
$(I)$ and $(II)$ transitions {\bf cannot} be used for any $c(*,*)$ on the northern border.

In particular, since only type $(III)-(VII)$ transitions can be used, all the lines of $w$ except the first
two, are composed of $\$$. \eproof

In order to obtain a successful ``encoding'' in $w$ of a solution for $PCP(x,y)$ we must prove that $k=r$ and
indexes $i_l=j_l$ $\forall l\in \overline{1,k}$. 

Using type $(III)-(VII)$ transitions, a $p+2$ line marks the letters of a corresponding pair of tiles
$(x_{i_p},y_{i_p})$ in the PCP solution. Final states and a final class are obtained only after successful
reduction of these tiles. Additional $\$$ lines parsed with type $(III)$ transitions can be added to such
final positions. This behavior is captured by the following lemma.\vsp

\noi{\bf Lemma 2}
\label{lema2}
Let $w$ be a grid with $m$ lines and $q$ columns parsed by $S$ and let us use the notation $x_{i_1}, \ldots
x_{i_k},y_{j_1}, \ldots y_{j_r}$ as in Lemma 1\out{\ref{lema1}}. Then:
\begin {enumerate}
\item For all $p$ with $0\leq p\leq k$, the southern border of the $p+2$ line of $w$ can be written 
$c(a_{p,1},b_{p,1}), \ldots , c(a_{p,q},b_{p,q})$ 
and satisfies the following equalities (as lists over $\{0,1,\ldots,n\}$):\\
$(a_{p,1},\ldots ,a_{p,q}) = (0^{\alpha},i_{p+1}^{|x_{i_{p+1}}|}, \ldots ,i_k^{|x_{i_k}|})$\\
$(b_{p,1},\ldots ,b_{p,q}) = (0^{\beta}, j_{p+1}^{|y_{j_{p+1}}|}, \ldots ,j_r^{|y_{j_r}|})$
\bi\item[]
where $\alpha=\sum_{z=1}^{p}|x_{i_z}|$ and $\beta=\sum_{z=1}^{p}|y_{j_z}|.$\ei
Furthermore, $k=r$ and $i_p=j_p$ $\forall p\in \overline{1,k}$.
\item Any line of $w$ after the $k+2$ line, if any, is parsed using type $(III)$ transitions only. 
\item The eastern border $b_e$ is of the following type $b_e \in A\cdot A \cdot M(i_1,0,0)\cdots
  M(i_k,0,0)\cdot \{A\}^{\star}$
\end{enumerate}

\bproof \vsp

{\it (i)} Recall from the previous lemma that the two equalities of lists hold for $p=0$. We will prove the
equalities by induction over $p$.

We refer to the index sequences $(a_{p,1},\ldots ,a_{p,q})$ and $(b_{p,1},\ldots ,b_{p,q})$, as the first
stream and the second stream of specifying the southern border of the $p+2$ line of $w$ (or the northern
border of the $p+3$ line, if such a line exists).

First, some remarks on type $(VII)$ transitions. These transitions can be composed horizontally only with
transitions of the same type. These transitions are only defined if the first index of the classes $M$ on the
western and eastern border are equal. Therefore, whenever a class $M(i,k_1,k_2)$ appears on a line, only
$M(i,\_,\_)$ classes can appear when parsing the rest of the line. Furthermore, whenever $k_1=0$, the second
index in $M(i,\_,\_)$ classes remains $0$ until the end of the line and the rest of the first northern stream
is copied to the first southern stream. Similarly, whenever $k_2=0$ the third index in $M(i,\_,\_)$ classes
remains $0$ until the end of the line and the rest of the second northern stream is copied to the second
southern stream.

Let $p=1$. From the previous lemma, the northern border of the first letter in the third line is $c(i_1,j_1)$
with $i_1,j_1\geq 1$ and the western border is $A$. Then $i_1=j_1$ since only a type $(IV)$ transition can be
used to parse this letter and the transitions used to parse the rest of line 3 are only type $(VII)$
transitions, with $M(i_1,\_,\_)$ on the western and eastern borders. The processing of the following letters
on the third line is deterministic, since depending on the northern and western border, at most one type
$(VII)$ transition can be chosen at each step.

Let $\alpha=|x_{i_1}|\geq 1$ and $\beta=|y_{i_1}|\geq 1$ (recall that both lists in PCP have nonempty
words). The first letter of this line is processed using the following type $(IV)$ transition:
$$\cross{c(i_1,i_1)}{\addspace{.8cm}{A}}{\$}{M(i_1,\alpha-1,\beta-1)}{\addspace{.4cm}{c(0,0)}}$$

For the sake of simplicity, assume $\alpha < \beta$. Then the following $\alpha-1$ letters in this line are
parsed using type $(VII)$ transitions, each creating a southern border of $c(0,0)$ and decreasing the second
and third indexes in $M(i,\_,\_)$ by 1 until $k_1$ reaches 0 using
$$\cross{c(i_1,i_1)}{M(i_1,k_1,k_2)}{\$}{M(i_1,k_1-1,k_2-1)}{\addspace{.4cm}{c(0,0)}}.$$ The first and second
southern streams have $\alpha$ leading zeros and $k_1=0$ is carried over when parsing the rest of the
line. The first equality is thus proven since the rest of the first northern stream is copied to the first
southern stream.

After parsing $\alpha$ $i_1$'s from the first northern stream there are still at least
$\beta-\alpha=|y_{i_1}|-|x_{i_1}|$ $i_1$'s on the second northern stream. For the next $\beta-\alpha$ letters
only type $(VII)$ transitions can be used:
$$\cross{\addspace{.4cm}{c(j,i_1)}}{M(i_1,0,k_2)}{\$}{M(i_1,0,k_2-1)}{\addspace{.2cm}{c(j,0)}}$$ therefore the
second southern stream has a total of $\beta$ leading zeros. After that, $k_2$ reaches zero and the rest of
the second northern stream is copied to the second southern stream. Thus we have obtained the second required
equality and $M(i_1,0,0)$ on the easternmost border of line 3. A symmetrical argument holds when $\alpha\geq
\beta$.

Note that if $k>p\geq 1$ or $r>p\geq 1$, then the southern border of this line still contains non-final states
- $c(i_k,j_r)$ so $w$ has at least another line, which motivates the use of induction.

Let $p\in \overline{2,k}$. Let $$\alpha=\sum_{z=1}^{p-1}|x_{i_z}|\; and \;\beta=\sum_{z=1}^{p-1}|y_{j_z}|.$$
We distinguish only two non-similar cases: $\alpha=\beta$ or $\alpha<\beta$, due to the symmetry in type $(VII)$
transitions. 

If $\alpha=\beta$, then by the induction hypothesis, both northern streams of line $p+2$ have exactly $\alpha$
leading zeros followed by $q-\alpha$ nonzero numbers. This allows us to use $\alpha$ type $(III)$ transitions
to parse the first $\alpha$ letters on this line. The next letter has the northern border $c(i_p,j_p)$ which
forces a type $(IV)$ transition and therefore $i_p=j_p$. The two equalities are proved similarly with the case
$p=1$ above. Note that this case can be avoided entirely if we consider only atom solutions of $PCP(x,y)$
(i.e. no non-trivial prefix of the solution forms a valid solution).

If $\alpha<\beta$, then after the first $\alpha$ leading zeros, the first northern stream contains only
nonzero numbers, while the second one has another $\beta-\alpha$ leading zeros followed by $q-\beta$ nonzero
numbers. Then after $\alpha$ type $(III)$ transitions, the $\alpha+1$ atom has $c(i_p,0)$ on the northern
border and $A$ on the western border. The type $(V)$ transition used to parse the first letter is:
$$\cross{c(i_p,0)}{\addspace{.4cm}{A}}{\$}{M(i_p,\varphi-1,\psi)}{\addspace{.2cm}{c(0,0)}}.$$
where $\varphi=|x_{i_p}|$ and $\psi=|y_{i_p}|$.

The rest of the first northern stream contains at least $\varphi -1$ $i_p$'s that require parsing by type
$(VII)$ transitions with $k_1>0$. Thus, the corresponding first southern stream will contain an equal number
of zeros in that part of the stream. After that, only type $(VII)$ transitions with $k_1=0$ can be used which
copy the rest of the first northern stream to the first southern stream.

The rest of the second northern stream has another $\beta-\alpha-1$ leading zeros which force type $(VII)$
transitions with $k_2>0,\;j_2=0$, that copy the zeros to the second southern stream. Afterwards, the second
northern stream reaches the $\psi$ $j_p$'s created in line 2 by the word $y_{j_p}$ and carried over to this
line.

Since the case $k_2>0,\;j_p>0\;j_p\not = i_p$ is undefined for type $(VII)$ transitions, the only remaining
possibility is that $j_p=i_p$. Thus the next $\psi$ numbers from the second northern stream are replaced by
zeros in the second southern stream, decreasing $k_2$ until it becomes 0. Afterwards the rest of the second
stream is copied from north to south, yielding the eastern border $M(i_p,0,0)$.

This proves the required equalities. Moreover, $w$ must have an additional line and the process continues if
$p<min(k,r)$ because $k>p$ or $r>p$ imply that the southern border of the current line contains the non-final
state $c(i_k,j_r)$.

Assume $k<r$, then the southern border of line $k+2$ has a first stream containing only zeros, and a second
one with $\beta$ leading zeros followed by $0<q-\beta$ nonzero numbers. In this case, the southern border
contains the non-final state $c(0,j_r)$, therefore $w$ must have at least another line. However, in this case,
it is impossible to obtain a final class on the additional line. Indeed, the first $\beta$ pairs of zeros are
deterministically parsed by type $(III)$ transitions. The following $\beta+1$ letter, with $c(0,j_{k+1})$
on the northern border and $A$ on the eastern border, can only be parsed by a type $(VI)$ transition with
$M(j_{k+1},|x_{j_{k+1}}|,|y_{j_{k+1}}|-1)$ on the eastern border.

However, since the rest of the first northern border contains only zeros, the rest of the line can be parsed
only using type $(VII)$ transitions with $k_1=|x_{j_{k+1}}|>0$ and $j_1=0$. Therefore $r_1=k_1>0$ and the
eastern border of this line is $M(j_{k+1},|x_{j_{k+1}}|,0)$ which is not a final class! A similar argument for
$k>r$ proves that $k=r$.\vsp

{\it (ii \&\ iii)} If $m \geq k+3$, the northern border of the $k+3$ - th line of $w$ has both streams full of
0's. Therefore, only type $(III)$ transitions are used to parse this line, copying the northern border to the
next line and resulting in an eastern border of $A$. Similarly, any line of $w$ after the $k+2$ line has only
type $(III)$ transitions yielding $A$ on the eastern border. From Lemma 1\out{\ref{lema1}} the first two lines
have $A$ on the eastern border. From {\it (i)} the eastern border of the next $k$ lines is $M(i_1,0,0)\ldots
M(i_k,0,0)$ which concludes the proof.  \eproof

Using these two lemmas, one can prove the following result:\vsp\vsp

\noi{\bf Theorem 3}
The finiteness and the emptiness problems for finite interactive systems are undecidable.
\vsp

\bproof Assume that the emptiness problem is decidable. Then for each PCP instance, $PCP(x,y)$, we can use the
above construction in order to obtain the FIS $S=S(x,y)$. For this FIS, we can decide whether or not $L(S)$ is
empty. The previous lemmas show that for every $w \in L(S)$, the first two lines of $w$ determine a solution
for $PCP(x,y)$ and that from any solution of $PCP(x,y)$, a grid $w$ constructed from the strings of the
solution and composed vertically with enough $\$$'s ($k$ lines where $k$ is the length of the solution), can
be parsed according to the lemmas (therefore $w\in L(S)$). Therefore the assumption contradicts the fact that
PCP is undecidable.

For the undecidability of the finiteness problem, it suffices to observe that if $L(S)$ is not empty, then it
is infinite. That follows from the fact that if $w\in L(S)$ then $w\cdot \$^{\dagger}\in L(S)$ (parsing the
additional lines with type $(III)$ transitions). Also, the iteration (any number of times) of a solution for
$PCP(x,y)$ gives another $PCP(x,y)$ solution for which a $w' \in L(S)$ can be constructed by horizontal
iteration of $w$. Therefore $L(S)$ is empty iff it is finite which concludes the proof.  \eproof

\vspace{18pt}{\large \bf 4.~The accessibility of a given FIS transition
}\vspace{12pt}

For any instance of PCP, we construct a FIS $S1$ which has all the states, classes, and transitions of the FIS
$S$ in the previous section, an additional state $q$, and two additional classes $Q$,$T$.  The initial states
and classes of $S1$ are those of $S$, the only final state is $q$ and the only final class is $T$. In
addition to $S$, $S1$ has the following new transitions:
\begin{enumerate}
\item \cross{\addspace{.4cm}{s}}{\addspace{.75cm}{A}}{\$}{\addspace{.75cm}{T}}{\addspace{.1cm}{s}}
\item \cross{\addspace{.4cm}{s}}{\addspace{.15cm}{M(i,0,0)}}{\$}{\addspace{.75cm}{T}}{\addspace{.1cm}{s}} 
$\forall i\in \overline{1,n}$
\item \cross{c(0,0)}{\addspace{.75cm}{A}}{\$}{\addspace{.75cm}{Q}}{\addspace{.1cm}{q}}
\item \cross{c(0,0)}{\addspace{.75cm}{Q}}{\$}{\addspace{.75cm}{Q}}{\addspace{.1cm}{q}}
\item \cross{\addspace{.4cm}{s}}{\addspace{.75cm}{Q}}{\$}{\addspace{.75cm}{T}}{\addspace{.1cm}{q}}
\end{enumerate}

This construction may be used to get the following result:\vsp\vsp

\noi{\bf Theorem 4}
The accessibility of a transition in a FIS is undecidable.
\vsp

\bproof It suffices to prove that transition of type $(v)$ is accessible iff $PCP(x,y)$ has a solution.

If $PCP(x,y)$ has a solution then there is a $w$ with $m$ lines and $q$ columns accepted by the FIS in the
previous section. Because of the border conditions established by Lemma 1\out{\ref{lema1}} and
2\out{\ref{lema2}}, we can add one column to the right of $w$ and an additional line, obtaining $u\in
L(S1)$. Indeed if we take a successful running of $w$ in the previous FIS, and parse the last line of $u$ using
$q$ type $(iii)$ and $(iv)$ transitions, the last column with type $(i)$ and $(ii)$ transitions, and finally a
type $(v)$ transition for the southeasternmost letter of $u$, we obtain a successful running of $u$ on $S1$,
which uses the transition $(v)$.

If transition $(v)$ is accessible, let $u$ be a grid that can be parsed such that transition $(v)$ can be
applied to the southeasternmost letter. We shall prove that there exists $w$ such that $u = (w\triangleright
\$^{\star})\cdot \$^{\dagger}$ and $w$ is recognized by the FIS S in the previous section.

Firstly, the letter parsed with this type $(v)$ transition cannot be on the first column since $Q$ is not an
initial class. The only way of obtaining class $Q$ on the western border on this line is by using a type
$(iii)$ transition and some (if any) type $(iv)$ transitions. This implies that $c(0,0)$ is on the northern
border of all the atoms in the last line of $u$ and since $c(0,0)$ is not an initial state, $u$ has at least 2
lines. Similarly, the state $s$ cannot appear on the northern border of the last atom, unless only type $(i)$
and some (if any) type $(ii)$ transitions were used for processing the atoms on the last column of $u$. Then
there exists $w$ such that $u = (w\triangleright \$^{\star})\cdot \$^{\dagger}$. The southern border of $w$ is
composed only of $c(0,0)$'s and the eastern border of $A$'s and $M(i,0,0)$'s for some $i \in \overline{1,n}$.

Since type $(i)$-$(v)$ transitions cannot lead to $A$'s or $M(i,0,0)$'s on the eastern border, it follows that
$w$ was parsed using transitions of the FIS in the previous section. Furthermore, due to the southern and
eastern border conditions for $w$, this parsing is a successful running of $w$ in the FIS in the previous
section and therefore $PCP(x,y)$ has a solution.  \eproof

\vspace{18pt}{\large \bf 5.~The equivalence of FISs and tile systems
}\vspace{12pt}

In this section we present a simple, direct proof of the equivalence between FISs and tile systems. Tile
systems \cite{gi-re97,lmn98} are but one of many equivalent mechanisms specifying the class of recognizable
two-dimensional languages.

\paragraph{Tile systems}

For a grid $w$ of size $m\times n$, let $\widehat{w}$ denote the grid of size $(m+2)\times (n+2)$ obtained
bordering $w$ with a special symbol $\sharp$. A tile system is defined as follows:\bi
\item for a bordered grid $\widehat{w}$, let $B_{r,s}(\widehat{w})$ be the set of its sub-grids of size
  $r\times s$;
\item a two-dimensional language $L$ over $V$ is {\em local} if there is a set $\Delta$ of $2\times 2$ grids
  over $V\cup\{\sharp\}$ such that $$L=\{w\ |\ w\mbox{ grid over }V\wedge B_{2,2}(\widehat{w})\subseteq\Delta\}$$
\item a two-dimensional language $L$ over an alphabet $V$ is recognized by a {\em tile system} if there is an
  alphabet $V'$, a local language $L'$ over $V'$ and a letter-to-letter homomorphism $h:V'\ra V$ such that
  $h(L')=L$.  \ei

\paragraph{From FISs to tile systems}

Let $L$ be a grid language over an alphabet $V$ recognized by a FIS $F$. Let $V_1$ denote the extended
alphabet consisting of tuples $(N,W,a,E,S)$ ($N/W/E/S$ stands for north/west/east/south, respectively), with
$W,E$ classes in $F$, with $N,S$ states in $F$, with $a$ in $V$, and such that $\scruce{N}{W}{a}{E}{S}$ is a
transition in $F$.

Consider the local language $L_1$ over $V_1$ generated by the following set of $2\times 2$ tiles over
$V_1\cup\{\sharp\}$:\bi
\item (tiles with letters in $V_1$ which agree on the connecting
  cells)\\ $\begin{array}{c@{}c}w_1&w_2\\w_3&w_4\end{array}$, with $w_k=(N_k,W_k,a_k,E_k,S_k), k\in\ol{1,4}$
    and such that $E_1=W_2, S_1=N_3, S_2=N_4, E_3=W_4$.
\item (tiles with $\sharp$ for handling the borders) - these are tiles with $\sharp$ and such that the next
  elements in $V_1$ have appropriate initial/final states/classes; \\--~for instance, the north-east corner
  tile is $\begin{array}{c@{}c}\sharp&\sharp\\\sharp& w_4\end{array}$ with $N_4,W_4$ initial; \\--~the
    middle-north border tile is $\begin{array}{c@{}c}\sharp&\sharp\\w_3&w_4\end{array}$ with $N_3,N_4$
      initial; \\--~and so on.\ei

One can easily see that the grids in $L_1$ correspond to the scenarios in $F$. By dropping the information
around the transition symbols with the homomorphism $h: (N,W,a,E,S)\mapsto a$ one gets a tile systems
specifying $L$. This proves one implication, namely:\vsp

\noi{\bf Lemma 5}
A language recognized by a FIS can be specified with a tile system.
\vsp

\paragraph{From tile systems to FISs}

It is obvious that FIS languages are closed to letter-to-letter homomorphism, so we can restrict ourself to
local languages.

For a local language $L$ over an alphabet $V$ and specified by a set $\Delta$ of tiles, we construct an
equivalent FIS $F$ as follows:\bi
\item the set of states is the same as the set of classes and consists of $2\times 2$ tiles over
  $V\cup\{\sharp\}$ from $\Delta$;
\item the transitions are
  $\scruce{\begin{array}{c@{}c}N_1&N_2\\N_3&N_4\end{array}}{\begin{array}{c@{}c}W_1&W_2\\W_3&W_4\end{array}}
{a}{\begin{array}{c@{}c}E_1&E_2\\E_3&E_4\end{array}}{\begin{array}{c@{}c}S_1&S_2\\S_3&S_4\end{array}}$ with:
\\--~$(N_1,N_2,N_3,N_4)=(W_1,W_2,W_3,W_4)$, 
\\--~$N_4=W_4=a$, 
\\--~$(N_3,N_4)=(S_1,S_2)$, and 
\\--~$(W_2,W_4)=(E_1,E_3)$; 
\item initial states are tiles $\begin{array}{c@{}c}\sharp&\sharp\\N_3&N_4\end{array}\in\Delta$;\\ initial
  classes are tiles $\begin{array}{c@{}c}\sharp\ &W_2\\\sharp\ &W_4\end{array}\in\Delta$; 
\item final states are tiles $\begin{array}{c@{}c}N_1&N_2\\\sharp&\sharp\end{array}\in\Delta$;\\ final classes
  are tiles $\begin{array}{c@{}c}W_2&\ \sharp\\W_4&\ \sharp\end{array}\in\Delta$; \ei 

A grid $w$ is in $L$ if and only if it is recognized by $F$. In this way, the following reverse implication
result is proved:\vsp

\noi{\bf Lemma 6}
A language specified by a tile system can be recognized with a FIS.
\vsp

The two lemmas above prove the main result of this section.\vsp

\noi{\bf Theorem 7}
A language is recognized by a FIS if and only if it can be specified with a tile system.
\vsp

Theorem~7\out{\ref{th-fis-ts}} and a result in \cite{la-si97b}, stating that by extracting the first line from
the grids specified by tile systems one gets precisely the context-sensitive string languages, yield an
alternative, indirect proof of the undecidability of the emptiness problem for FISs.

\vspace{18pt}{\large \bf 6.~Conclusions and future work
}\vspace{12pt}

We have proved that a few simple and easily decidable properties on finite automata (like accessibility of a
transition, finiteness, etc.) become undecidable when extended to finite interactive systems. It may be
worthwhile to find interesting restricted classes of FIS languages for which these properties are decidable.

One can also look at possible extensions of the technique in this paper for covering different sets of
restricted grids such as: non-rectangular grids, connected grids, bounded grids, etc.  For these classes, one
may use the {\it bounded Post Correspondence Problem} (which bounds the number of pairs used in a PCP solution
to no more than $k$, including repeated tiles) which is also known to be NP-complete \cite{guide79}.

Other open area is to develop an algebraic theory for representing FIS languages, similar to the regular
algebra used for regular languages and finite automata.

\vspace{18pt}{\large \bf Acknowledgment
}\vspace{12pt}

This research was partially supported by the GlobalComp Grant (PNCDI-II, Project 11052/18.09.2007, Romania).

\end{document}